\documentclass[letterpaper,12pt]{article}
\usepackage{graphics}
\newcommand{\tv}[1]{\mathbf{#1}}
\newcommand{\XX}{\tv{X}}
\newcommand{\DD}{\tv{\Delta}}
\newcommand{\xx}{\tv{x}}
\newcommand{\xmin}{x_{\mathrm min}}
\newcommand{\xmax}{x_{\mathrm max}}

\begin{document}
\section*{Data analysis recipes:\ \\
  Choosing the binning for a histogram\footnote{%
    Copyright 2008 David~W.~Hogg (david.hogg@nyu.edu).  You may copy
    and distribute this document provided that you make no changes to
    it whatsoever.}}

\noindent
David~W.~Hogg\\
\textsl{Center~for~Cosmology~and~Particle~Physics, Department~of~Physics\\
New~York~University}\\
\texttt{david.hogg@nyu.edu}

\begin{abstract}
  Data points are placed in bins when a histogram is created, but
  there is always a decision to be made about the number or width of
  the bins.  This decision is often made arbitrarily or subjectively,
  but it need not be.  A jackknife or leave-one-out cross-validation
  likelihood is defined and employed as a scalar objective function
  for optimization of the locations and widths of the bins.  The
  objective is justified as being related to the histogram's
  usefulness for predicting future data.  The method works for data or
  histograms of any dimensionality.
\end{abstract}

\section{Introduction}

There are many situations in experimental science in which one is
presented with a collection of discrete measurements $\xx_j$ and one
must bin those points into a set of finite-sized bins $i$, with
centers $\XX_i$ and full-widths $\DD_i$, to create a histogram of
numbers of points $N_i$, or the equivalent when the points have
non-uniform weights $w_j$.  The problem of binning comes up, for example,
when one needs to plot a data histogram, when one needs to perform
least-square fitting of a probability distribution function, and when
one wants to compute entropies or other measurements on the inferred
data probability distribution function.

The choice of bin centers and widths often seems arbitrary.  However,
there is a non-arbitrary choice, derived below, which emerges when the
histogram is thought of as an estimate of the probability distribution
function of whatever process generated the data.  If the binning is
too coarse, the histogram does not give much information about the
shape of the probability distribution function.  If the binning is too
fine, bins become empty and the histogram becomes noisy, so it in some
sense ``overfits'' the data.  The best binning lies in between these
extremes and can be found simply and quickly by a ``jackknife'' or
cross-validation method, that is, by excluding data subsamples and
using the non-excluded data to predict the excluded data.  This is not
the only data-based binning-choice approach\footnote{\raggedright see,
  for example, Knuth,~K.~H., ``Optimal data-based binning for
  histograms,'' arXiv:physics/0605197, and references cited therein.},
but it is simple and sensible.

In what follows, we are going to consider a data histogram, which we
imagine as a set of bins $i$, with centers $\XX_i$ and widths (or
multi-dimensional volumes) $\DD_i$.  Equivalently (and perhaps more
usefully), the parameterization of the bins can be described by a set
of edges $\XX_{(i-1/2)}$ so the centers become
$\XX_i=\left(\XX_{(i-1/2)}+\XX_{(i+1/2)}\right)/2$ and the widths
become $\DD_i=\left|\XX_{(i+1/2)}-\XX_{(i-1/2)}\right|$.  These bins
will get filled by a set of (possibly multi-dimensional) data points
$\xx_j$, leading to each bin $i$ containing a number of data points
$N_i$.  We will also make reference to the binning function $i(\xx)$
which, for a given data value $\xx$, returns the bin $i$.

\section{Model probability distribution function}

Our best binning is based on the idea that the histogram is a sampling
of a probability distribution function and can therefore be thought of
as providing an estimate or model of that probability distribution
function.

One possible (approximate) probabilistic model for the data is that
they are drawn from a probability distribution function such that, in
each bin of the histogram we are making, the probability is constant
and proportional to the number of actual data points that landed (by
chance) in that bin.  This model has the limitation that bins that
happen (by chance) to be empty will be assigned zero probability; when
a new datum happens to arrive (by chance) inside one of those
previously empty bins, it will be assigned a vanishing likelihood and
render the probabilistic model false at (arbitrarily) high confidence.

A more well-behaved (approximate) probabilistic model is that the
probability $p(i)$ that a data point land in bin $i$ is
\begin{equation}
p(i) = \frac{N_i+\alpha}{\displaystyle\sum_k\left[N_k+\alpha\right]}\quad,
\end{equation}
where $\alpha$ is a dimensionless ``smoothing'' constant of order
unity (to be set later).  Here, so long as there are a finite number
of bins, the probability is non-zero in every bin.  The associated
(approximate, model) probability distribution function is
\begin{equation}
\tilde{f}(\xx) = \frac{p\left(i\left(\xx\right)\right)}{\DD_{i(\xx)}}\quad,
\end{equation}
where $i(\xx)$ is the function that returns the bin $i$ for any value
$\xx$.  Note that the function $\tilde{f}(\xx)$ is normalized by
construction;
\begin{equation}
\int\tilde{f}(\xx)\,\mathrm{d}\xx = 1 \quad.
\end{equation}

In general, the data points will not all be treated equally, but in
fact each data point $\xx_j$ will come with a weight $w_j$, and each
bin $i$ will contain total weight $W_i$.  The only change this makes
is in the inferred probability $p(i)$, which becomes
\begin{equation}
p(i) = \frac{W_i+\alpha}{\displaystyle\sum_k\left[W_k+\alpha\right]}\quad,
\end{equation}
where now the smoothing constant $\alpha$ will be of order the mean
weight $w_j$.

\section{Jackknife likelihood}

Imagine now that a new datum is recorded and happens to fall in bin
$i$.  The (logarithmic) likelihood of this new datum (according to the
approximate model) is just $\ln\tilde{f}(\xx)$.  If the binning is
extremely fine ($\DD_i$ small), then most bins will be empty and
assigned identical probabilities.  If the binning is extremely coarse
($\DD_i$ large), then although most bins will have high probabilities,
they will not have large values of $\tilde{f}(\xx)$ because they will
have large widths.  In either case, the predictive power of the model
probability distribution function $\tilde{f}(\xx)$ is low.  For most
well-behaved continuous (true) probability distribution functions,
there is a finite binning at which the likelihoods of new data are
maximized.

With a finite data set, a ``jackknife'' or leave-one-out
cross-validation likelihood $L$ can be defined to be the total
weighted (logarithmic) likelihood of each data point $\xx_j$ as
computed from the model probability distribution function
$\tilde{f}(\xx)$ computed from all the data points \emph{other than}
point $j$.
\begin{equation}
L = \sum_j w_j\,\ln\left(
  \frac{W_{i(\xx_j)}+\alpha-w_j}{\displaystyle
        \DD_{i(\xx_j)}\,\left[\sum_k\left[W_k+\alpha\right]-w_j\right]}
\right)\quad,
\end{equation}
where, again, $i(\xx_j)$ is the function that returns the bin $i$
containing the data point $\xx_j$.

In the simple case of no weighting (or, equivalently, $w_i=1$ for all
$i$), this jackknife likelihood can be written as
\begin{equation}
L = \sum_i N_i\,\ln\left(
  \frac{N_i+\alpha-1}{\displaystyle
        \DD_i\,\left[\sum_k\left[N_k+\alpha\right]-1\right]}
\right)\quad,
\end{equation}
where the sum over data points has been converted into a sum over
bins, because the latter is generally far faster.

The ``best'' binning parameters $\XX_i$, $\DD_i$ and $\alpha$ are
those that maximize the jackknife likelihood $L$.  This defines a
non-arbitrary choice of binning.  The choice is also motivated; it is
the choice that best predicts future data, under the assumption that
the existing data are representative.

\section{Examples}

\begin{figure}
\resizebox{\textwidth}{!}{\includegraphics{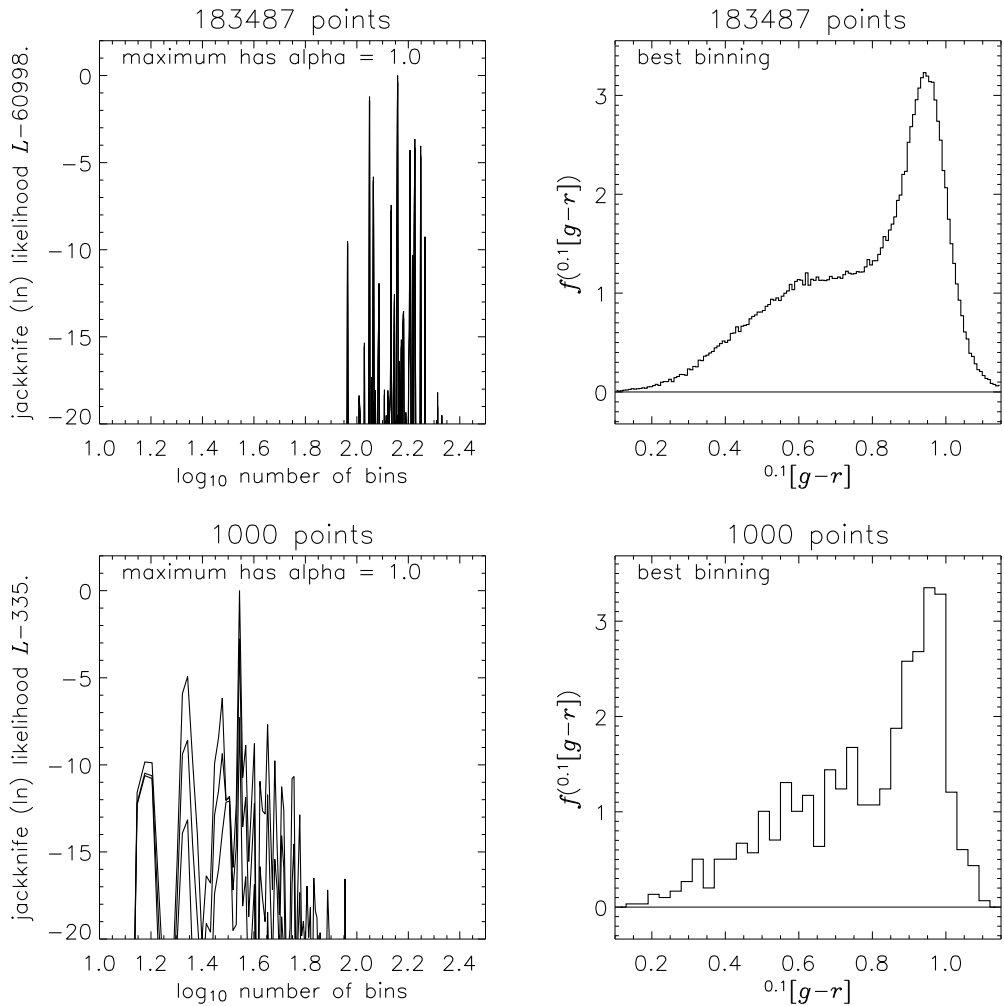}}
\caption[1d]{Constant-width binning of a set of measured galaxy colors.
The top-left panel shows grid searches in binsize for the eight
possible combinations of smoothing $\alpha= (10, 1, 0.1, 0.01)$ and
binning phase $\delta= (0, 0.5)$ (see text for definitions).  The
top-right panel shows the data binned with the maximum-likelihood
binning parameters.  The bottom panels show the same, but for a
randomly chosen subsample.\label{fig:binning_1d}}
\end{figure}

As a simple test, consider equal-width (all $\Delta_i$ equal) binnings of
a set of (one-dimensional) measurements $x_j$ (galaxy colors in this
case), in a fixed range $\xmin < x < \xmax$.  In this simple
situation, the binning only has two parameters: the number $N$ of bins
(which, given the color range, fixes the bin positions $\XX_i$ and
common widths $\DD_i\equiv\DD$) and the smoothing $\alpha$.  The
binning function $i(x)$ is then simply
\begin{equation}
i(x) = \mathrm{floor}\left(\frac{x-\xmin}{\Delta}
                           -\delta\right)\quad,
\end{equation}
where
\begin{equation}
\Delta\equiv \frac{\xmax-\xmin}{N}\quad.
\end{equation}
Figure~\ref{fig:binning_1d} shows the results of a grid search in this
parameter space for the optimum binning for the $^{0.1}[g-r]$ colors
of a large number of galaxies, and the same for a smaller subsample.

\begin{figure}
\resizebox{\textwidth}{!}{\includegraphics{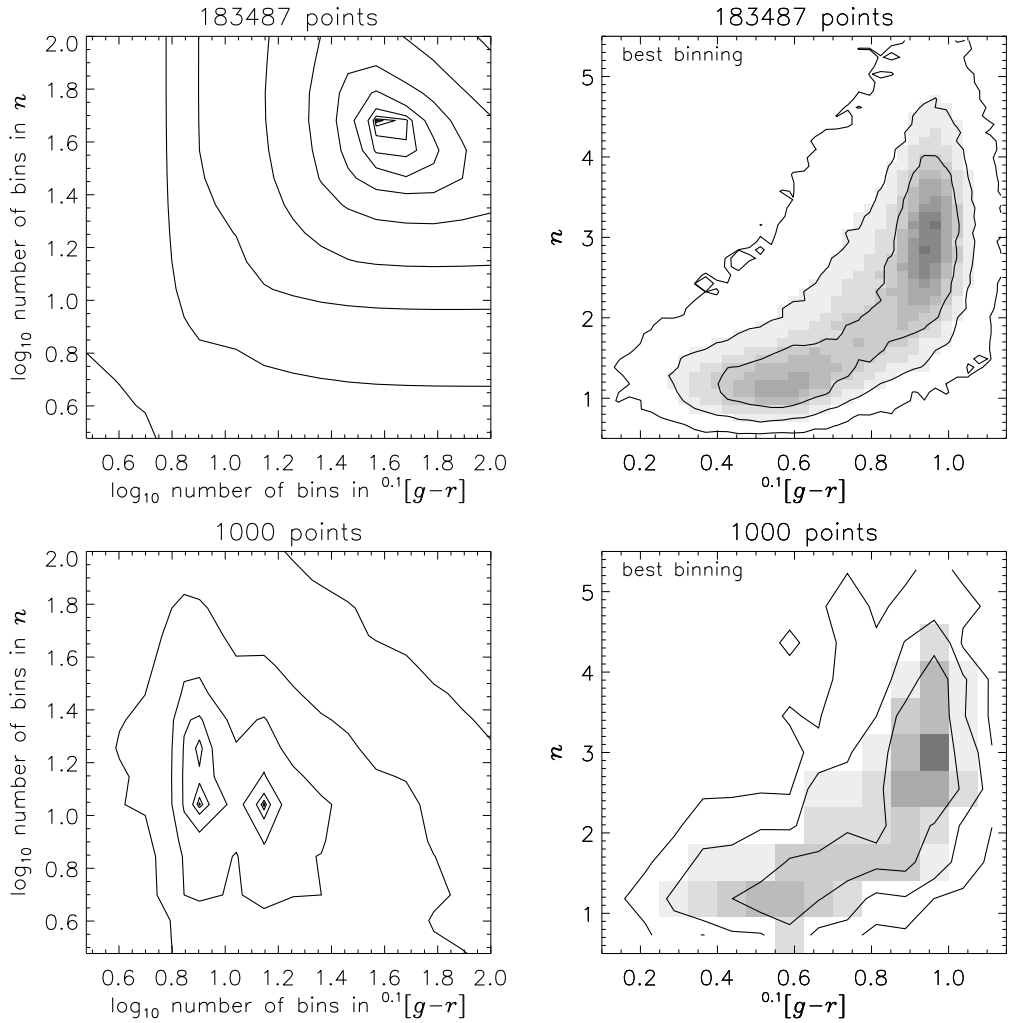}}
\caption[2d]{Two-dimensional constant-width binning of a set of measured
galaxy colors and radial profile shapes (as parameterized by the
S\'ersic index $n$).  The top-left panel shows a grid search in the
two binsizes, with smoothing fixed at $\alpha=1.0$ and both phases
fixed at $\delta=0$.  The top-right panel shows the data binned with
the maximum-likelihood binning parameters, plus contours at 2, 10, 25,
50, and 75~percent of the maximum value.  The bottom panels show the
same, but for a randomly chosen subsample.\label{fig:binning_2d}}
\end{figure}

There is nothing special (except simplicity) about the one-dimensional
case.  Figure~\ref{fig:binning_2d} shows the results of a grid search
for the optimal two-dimensional equal-width binning for two quantities
(the $^{0.1}[g-r]$ colors and S\'ersic indices $n$ of the same set of
galaxies).  In the two-dimensional case, the optimal binning is
coarser (because the space is ``bigger'').

\section{Discussion}

I have shown that when a histogram of data needs to be made, there
\emph{is} a non-arbitrary choice of binning.  Some qualitative
observations follow.
\begin{itemize}
\item
The optimal bin widths get smaller as the number of data points goes
up or as the features in the (true) probability distribution function
get narrower.
\item
The results are more sensitive to the smoothing parameter $\alpha$
when the number of empty or near-empty bins becomes significant.
\item
The jackknife likelihood makes discontinuous jumps as the bin edges
cross individual data points.  For this reason, the likelihood does
not have well-defined derivatives.  Some care must be taken that the
optimization technique does not depend on having a differentiable
likelihood function.
\item
There is nothing special about one-dimensional or two-dimensional
distribution functions; this is easily generalized to $n$-dimensional
distributions.  However, it takes a lot of data points to measure a
distribution function in $n$ dimensions when $n$ is large; I
understand that the required number of data points scales worse than
$e^n$ [need ref].
\item
There is nothing special about equal-width binning; I simply chose
this to make the optimization problem easily tractable and the results
easily presentable.
\item
This method makes no reference to the \emph{errors} or
\emph{uncertainties} on the measurements $\xx_j$.  Effectively, I have
assumed that the errors are small relative to any real features in the
probability distribution function.  In practice, it is rarely useful
to have more than a few bins per the width of your error distribution,
if all the points have similar uncertainties.
\item
There is often an additional choice about what minimum and maximum
data values to allow for histogramming.  This choice also ought to be
made in a non-arbitrary fashion if there are data points that will be
excluded by the choice.
\item
Finally, there is nothing special about the ``tophat'' binning model
used in the above examples.  Everything can be generalized to smoothly
overlapping bins, in which points are assigned fractionally to
multiple bins.  In general, smoother binning models make for more
well-behaved derivatives of the jackknife likelihood and therefore
more straightforward optimization.  This can also all be generalized
to kernel-smoothing techniques for density estimation, which ought to
be made the subject of a separate note.
\end{itemize}

\paragraph{acknowledgments}
It is a pleasure to thank Sam Roweis for useful discussions and
Michael Joyce and the Laboratoire de Physique Th\'eorique at
Universit\'e de Paris at Orsay for generous hospitality.

\clearpage

\clearpage

\end{document}